\documentclass{Interspeech2024}
\usepackage[acronym, shortcuts, nohypertypes={acronym}]{glossaries}
\usepackage{caption}
\usepackage{subcaption}
\usepackage{svg}
\usepackage[outdir=./]{epstopdf}

\usepackage[activate]{microtype}
\interspeechcameraready 

\title{Are you sure? Analysing Uncertainty Quantification Approaches for Real-world Speech Emotion Recognition}

\name[affiliation={1}]{Oliver}{Schrüfer}
\name[affiliation={2}]{Manuel}{Milling}
\name[affiliation={1}]{Felix}{Burkhardt}
\name[affiliation={1}]{Florian}{Eyben}
\name[affiliation={1}{2}{3}]{Björn}{Schuller}

\address{
  $^1$audEERING GmbH, Germany\\
  $^2$CHI -- Chair of Health Informatics, MRI, Technical University of Munich, Germany \\
  $^3$GLAM -- the Group on Language, Audio, \& Music, Imperial College London, London, UK
 }
\email{oschruefer@audeering.com, manuel.milling@tum.de, [fburkhardt, fe, bs]@audeering.com}

\newacronym{SER}{SER}{speech emotion recognition}
\newacronym{OOD}{OOD}{Out-of-Distribution}
\newacronym{EDL}{EDL}{Evidential Deep Learning}
\newacronym{emodb}{EmoDB}{Berlin Database of Emotional Speech}
\newacronym{cremad}{CREMA-D}{CRowd-sourced Emotional Mutimodal Actors Dataset}
\newacronym{ASR}{ASR}{automatic speech recognition}
\newacronym{AUC}{AUC}{area under the curve}
\newacronym{MC}{MC}{Monte Carlo}
\newacronym{ID}{ID}{in-domain}
\newacronym{DL}{DL}{Deep Learning}
\newacronym{UAR}{UAR}{Unweighted Average Recall}
\newacronym{SNR}{SNR}{Signal to Noise Ratio}
\newacronym{CDF}{CDF}{Empirical Cumulative Distribution Function}
\newacronym{uq}{UQ}{Uncertainty Quantification}
\newacronym{PCC}{PCC}{Pearson Correlation Coefficient}
\newacronym{PN}{PN}{Prior Networks}
\newacronym{SOTA}{SOTA}{State-of-the-art}
\newacronym{KL}{KL}{Kullback–Leibler}

\keywords{Uncertainty Quantification, Speech Emotion Recognition, Prior Networks, EDL, Out-of-Distribution}

\begin{document}

\maketitle


\begin{abstract}

\ac{uq} is an important building block for the reliable use of neural networks in real-world scenarios, as 
it
can be a useful tool in identifying faulty predictions.
Speech emotion recognition (SER) models can suffer from particularly many sources of uncertainty, such as the ambiguity of emotions, \ac{OOD} data or, in general, poor recording conditions. 
Reliable \ac{uq} methods are thus of particular interest as in many SER applications no prediction is better than a faulty prediction. 
While the effects of label ambiguity on uncertainty are well documented in the literature, we focus our work on an evaluation of \ac{uq} methods for SER under common challenges in real-world application, such as corrupted signals, and the absence of speech.
We show that simple \ac{uq} methods can already give an indication of the uncertainty of a prediction and that training with additional \ac{OOD} data can greatly improve the identification of such signals. 

\end{abstract}

\section{Introduction}\label{sec:intro}
Intelligent audio analysis is becoming more and more important in our lives, whether in voice assistant systems or assistive robots, e.\,g., in the care sector.
While the field of \ac{ASR} has seen major advances through \ac{DL}, enabling robust real-world applications, \ac{SER} still sees some major challenges after more than two decades of research \cite{schuller2018speech}.
A particular challenge hereby is the ambiguity, subjectivity \cite{sahu2019multimodal} or even untypical expression \cite{milling2022evaluating} of emotions and thus the lack of ground truth labels\cite{rizos2020average}.
Even though a rough intuition of emotions can be considered common sense, a wide range of emotion definitions and an almost arbitrary granularity of emotions make it already challenging to compare models between datasets \cite{gerczuk2021emonet}. 
A common approach to mitigate these challenges, which we also apply in this work, is to model only a subset of the available emotions, which tend to be available in most datasets and have a high agreement amongst annotators. 

Still, robust cross-corpus \ac{SER} models are hard to develop and they struggle when confronted with \ac{OOD} data.  
Beyond this, the paralinguistic information necessary to detect emotions is easily corrupted by real-world challenges, such as poor recording conditions or background noise \cite{jaiswal2021best}.

In line with the subjectivity of emotions, many application scenarios of emotional context do not require a pin-point accurate estimation of emotion.
Instead, they rather follow the philosophy of no (or a neutral) emotion prediction being preferable to a faulty emotion prediction.
An expressive example of this is 
the frequent discussion of confusion matrices when analysing \ac{SER} performance, where for instance mistaking happiness for anger is considered worse than the confusion with neutral.
\ac{SER} systems could thus benefit vastly from knowing when an emotion prediction is accurate and when it is not.  
A key approach to put this idea into practice is the quantification of uncertainty. 

Uncertainties have been studied extensively in the literature and can usually be divided into two or three different categories \cite{abdar2021review, gawlikowski2023survey}:
aleatoric uncertainty, caused by uncertainty in the data, on the one hand and epistemic uncertainty (model uncertainty), as well as distributional uncertainty on the other hand.
In the context of \ac{SER}, aleatoric uncertainty arises for instance through the ambiguity of the emotions themselves or through corrupted data.
Epistemic uncertainty arises from the model itself and describes the inability of the model to precisely model the underlying distribution.
One common reason for this is a lack of variability in the training data \cite{hullermeier2021aleatoric}, 
for instance unbalanced gender or age distribution or a mismatch of the language between training data and real-world application. 
Distributional uncertainty arises, as the name suggests, from \ac{OOD} samples.
In the case of \ac{SER}, this could be emotions which were not included in the training set or recordings that do not contain speech at all, e.\,g., music, silence or ambient sounds.
This is also often referred to as Out-of-Domain samples \cite{tan2023single, liu2020simple},
which for the sake of simplicity we will summarise with the mathematical term Out-of-Distribution under the abbreviation OOD.

Popular methods aiming to quantify these effects of uncertainty on a sample level can generally be split into four different types: single deterministic methods, Bayesian methods, ensemble methods and test-time augmentation methods \cite{gawlikowski2023survey}. 
The latter three all rely on multiple forward passes through a network for an uncertainty estimation and thus add significant computational cost at inference time, making efficient deployment challenging.
Single deterministic methods, however, only add minimal computational overhead and are thus promising candidates for real-time real-world uncertainty estimation in \ac{SER} applications.
Apart from being used directly by an application,
\ac{uq} can also help to leverage more sophisticated training methods, such as active \cite{zhang2013active} or weakly-supervised learning. %

In this work, we therefore aim to investigate the behaviour of several common (mostly single deterministic) \ac{uq} methods under simulated challenges of real-world \ac{SER} scenarios, in particular when confronted with \ac{OOD} data and different noise levels.   

To our knowledge, this is the first systematic investigation to evaluate the most prominent \ac{uq} methods on multiple \ac{SER} datasets under realistic challenges of real-world application.

\section{Uncertainty Tests}\label{sec:test}
In order to obtain a good picture of our \ac{SER} models with respect to different contexts, we measure the models' behaviour with respect to four different uncertainty tests.

\subsection{Rater Agreement}
A rater agreement test is a straight-forward approach to evaluate aleatoric uncertainty in the context of \ac{SER} \cite{han2017hard}. 
It calculates the sample-level correlation between the uncertainty of the model and the level of agreement of different annotators during the labelling process.
We thus obtain an impression of whether the sources of uncertainty are similar for a \ac{SER} model and human annotators.


\subsection{Unknown Emotions}
For uncertainty quantification in classification tasks, one of the main questions is, how the \ac{uq} method reacts to data points from classes not included during the training, in general aiming for higher uncertainties for unknown classes.
Accordingly, we will test our \ac{SER} models for samples with emotion labels that are unknown to the model.
It is important to note, however, that the expressiveness of this test suffers from limitations in the \ac{SER} case, as the ambiguity and potential overlaps of emotions may not be properly accounted for in the datasets\cite{amir2015comparing}.


\subsection{Non Speech Data}
A more clearly interpretable test is to expose the model to \ac{OOD} data, which does not contain speech at all, such as “background music” or environmental sounds.
In this case, it would be desirable for the \ac{uq} measures to have high scores, in order to avoid incorrect predictions.
Most real-world \ac{SER} applications apply Voice Activity Detection to prevent this type of faulty predictions, thus adding additional computational resources.




\subsection{Corrupted Signals}

Our final test aims to simulate bad recording quality, one of the major challenges of real-world applications.
For that purpose, we augment data with varying levels of white noise or environmental sounds and investigate the interplay of model performance and uncertainty predictions for increasing levels of noise.



\section{Uncertainty Quantification}\label{sec:quant}
With respect to the above discussed tests, we analyse four \ac{uq} methods, three of which belong to the category of single deterministic \ac{uq} methods, with only \ac{MC} Dropout needing several forward passes through the model.

\subsection{Entropy}\label{sec:entropy}
One of the easiest ways to get insights about the uncertainty of a categorical task is to calculate the entropy \cite{shannon1948mathematical} over logits.  
The entropy is calculated as
\(  H = - \sum_ {i=1}^N p_i log(p_i)\)
for \(N\) classes, where \(p_i\) is the probability to predict class \(i\).
We consider this method our baseline as it can be calculated a posteriori for each classification model trained with a standard softmax layer and (weighted) cross-entropy loss. 


\subsection{MC Dropout}
Many \ac{SOTA} models like PANNs \cite{kong2020panns} or wav2vec2.0 \cite{baevski2020wav2vec} contain dropout layers to prevent overfitting during training.
This again enables an out-of-the-box use of our second baseline, \ac{MC} dropout \cite{gal2016dropout}, which works by reactivating the dropout layers during inference to obtain multiple outputs for the same input sample.
The \ac{uq} is then calculated as the variance over the different outputs. 
For classification tasks, it is alternatively possible to use the mean probability per class to calculate the entropy, as in section \ref{sec:entropy}. 

\subsection{Evidential Deep Learning}

\ac{EDL} is a more sophisticated approach of \ac{uq} and formulates learning as an evidence acquisition process \cite{sensoy2018evidential, Amini_2020}, where each of the training samples adds to an evidential distribution. 
It is based on the Dempster-Shafer theory of belief functions \cite{liu2008classic}.
For application, Sensoy et al.\ \cite{sensoy2018evidential} specify non-negative belief masses \(b_k\) for all $K$ classes and an uncertainty mass \(u\), which all together add up to one \(u + \sum_{k=1}^Kb_k = 1.\)
The model outputs $K$ values, describing the evidence per class \(e_k\).
The evidences are thereby modelled by a Dirichlet distribution with the parameters \(\alpha = (\alpha_1, .., \alpha_K)\), as \(\alpha_k = e_k + 1\), which are then used to calculate the uncertainty and belief masses.

\subsection{Prior Networks}

\begin{figure}[t!]
    \centering
    \includegraphics[width=0.98\linewidth]{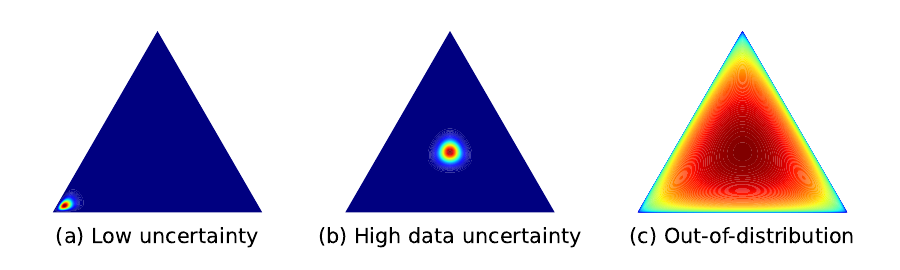}
    \caption{Simplexes of Dirichlet Distributions for a 3-class problem with sharp (a,b) and flat (c) distributions}
    \label{fig:pn}
\end{figure}
Another approach specifically tailored to detect \ac{OOD} samples are \ac{PN}s \cite{malinin2018predictive}.
Similar to \ac{EDL}, \ac{PN}s formulate the ground truth via an underlying Dirichlet distribution (see Figure \ref{fig:pn}).
This further allows for a distinction between data uncertainty and \ac{OOD} uncertainty.
By defining sharp distributions towards the respective classes, the network is trained for its classification task (Figure \ref{fig:pn} a)).
By specifying a flat distribution (Figure \ref{fig:pn} c)) it is possible to train on \ac{OOD} data without the need for an additional \ac{OOD} class.
For a 4-class-problem for instance, one would use, instead of the one hot encoding \((1,0,0,0)\), a Dirichlet distribution with \(\alpha = (101,1,1,1)\) and for \ac{OOD} data \(\alpha =(1,1,1,1)\) with 
the \ac{KL} divergence as the loss function.
For \ac{PN}s it is possible to differentiate between data uncertainty and \ac{OOD} data by calculating the precision (sharpness) of the distribution.
For the sake of simplicity, however, we will only refer to the combined uncertainty in this paper, which is once more calculated via the entropy.



\section{Experiments}\label{sec:exp}

\begin{figure*}[ht]

    \centering
    \begin{subfigure}[b]{0.19\textwidth}
    \caption*{Cross Entropy (CE)}
    \includegraphics[width= \linewidth]{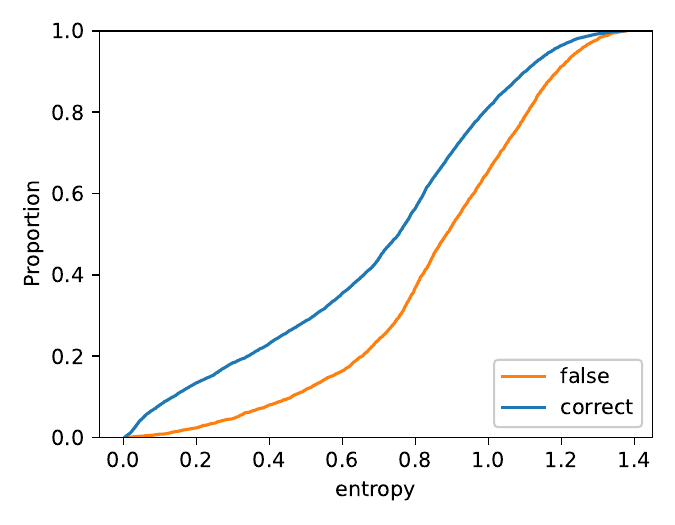}
    \end{subfigure}
    \begin{subfigure}[b]{0.19\textwidth}
    \caption*{MC Dropout}
    \includegraphics[width= \linewidth]{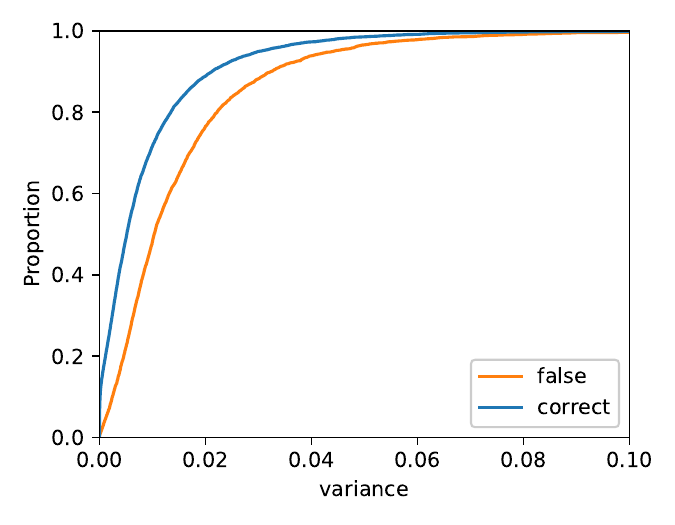}
    \end{subfigure}
    \begin{subfigure}[b]{0.19\textwidth}
    \caption*{Evidential DL}
    \includegraphics[width= \linewidth]{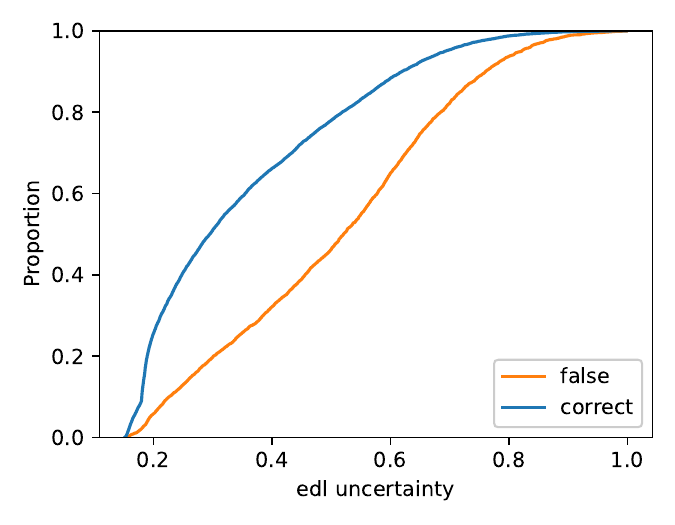}
    \end{subfigure}
    \begin{subfigure}[b]{0.19\textwidth}
    \caption*{Prior (in)}
    \includegraphics[width= \linewidth]{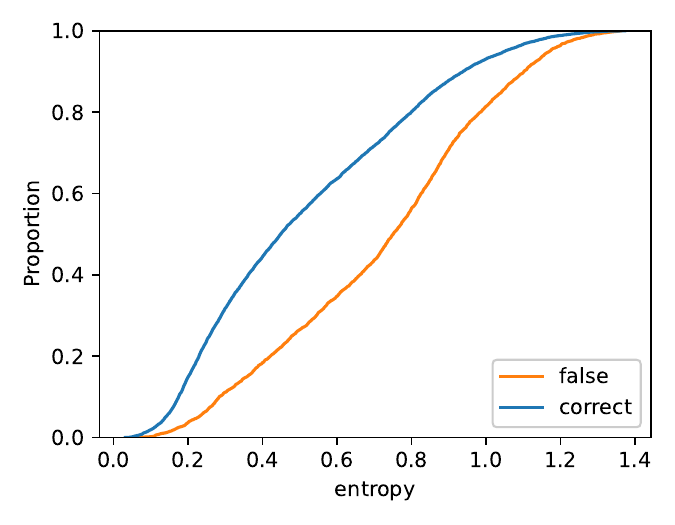}
    \end{subfigure}
    \begin{subfigure}[b]{0.19\textwidth}
    \caption*{Prior (out)}
    \includegraphics[width= \linewidth]{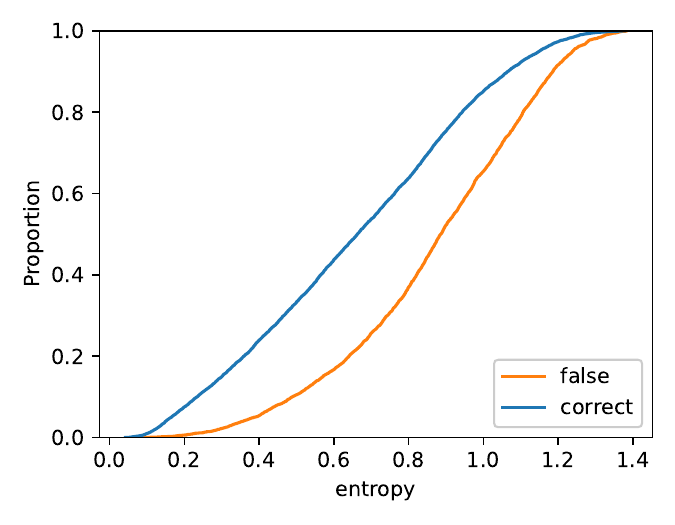}
    \end{subfigure}

    

    \begin{subfigure}[b]{0.99\textwidth}
    \includegraphics[width= \linewidth]{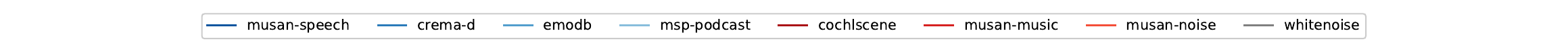}
    \end{subfigure}
    
    \begin{subfigure}[b]{0.19\textwidth}
    \includegraphics[width= \linewidth]{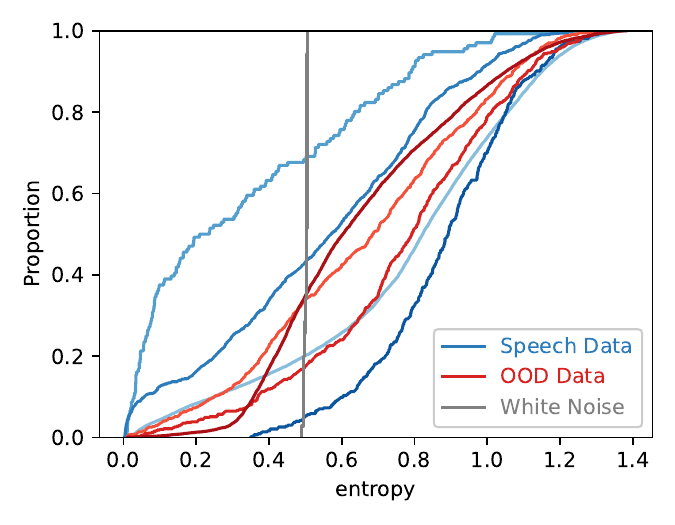}
    \end{subfigure}
    \begin{subfigure}[b]{0.19\textwidth}
    \includegraphics[width= \linewidth]{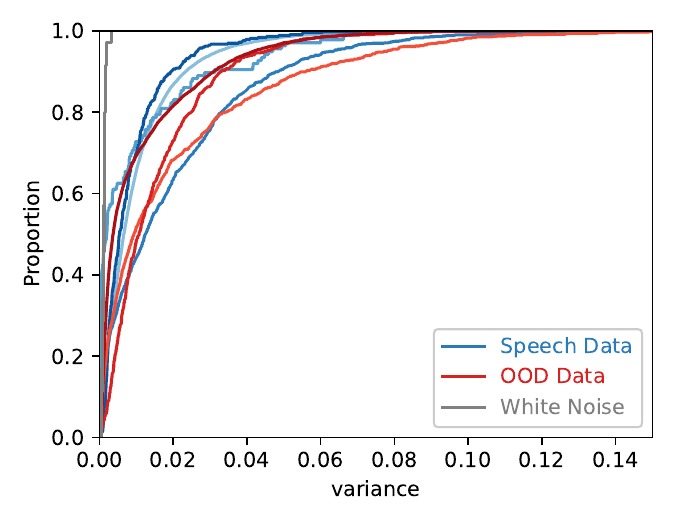}
    \end{subfigure}
    \begin{subfigure}[b]{0.19\textwidth}
    \includegraphics[width= \linewidth]{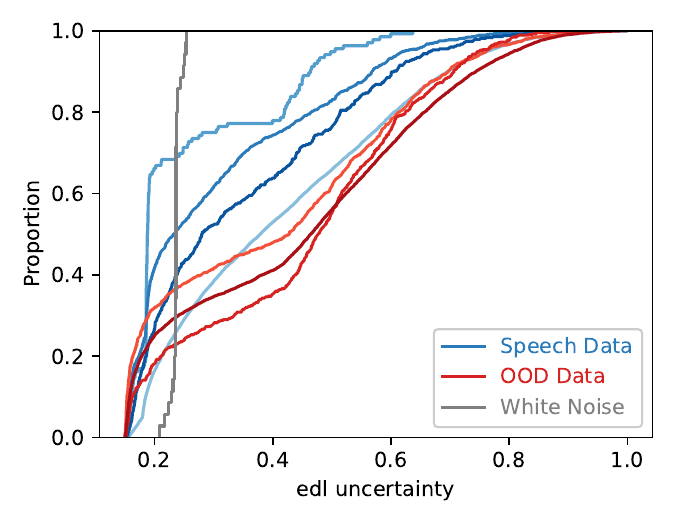}
    \end{subfigure}
    \begin{subfigure}[b]{0.19\textwidth}
    \includegraphics[width= \linewidth]{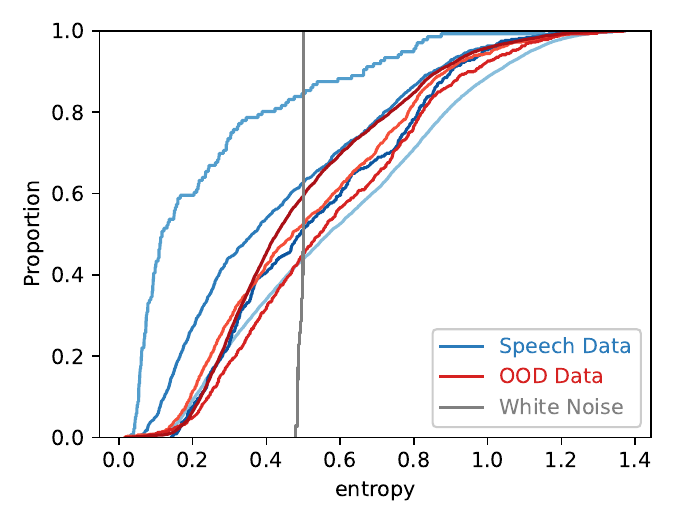}
    \end{subfigure}
    \begin{subfigure}[b]{0.19\textwidth}
    \includegraphics[width= \linewidth]{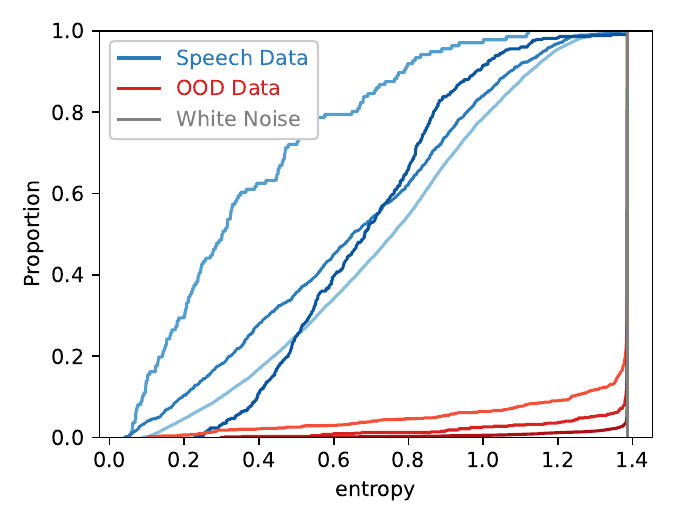}
    \end{subfigure}

    \begin{subfigure}[b]{0.19\textwidth}
    \includegraphics[width= \linewidth]{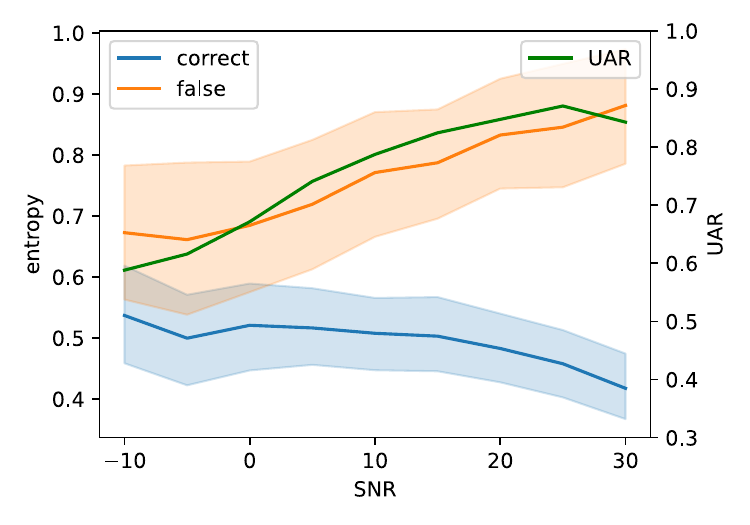}
    \caption*{Cross Entropy (CE)}
    \end{subfigure}
    \begin{subfigure}[b]{0.19\textwidth}
    \includegraphics[width= \linewidth]{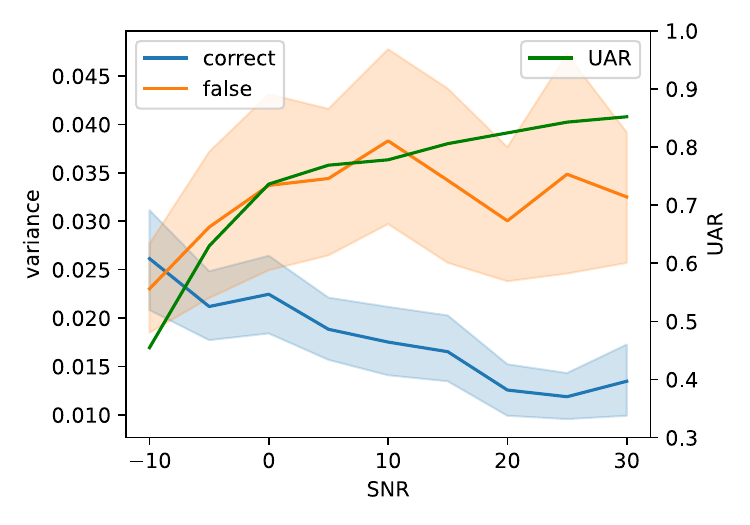}
    \caption*{MC Dropout}
    \end{subfigure}
    \begin{subfigure}[b]{0.19\textwidth}
    \includegraphics[width= \linewidth]{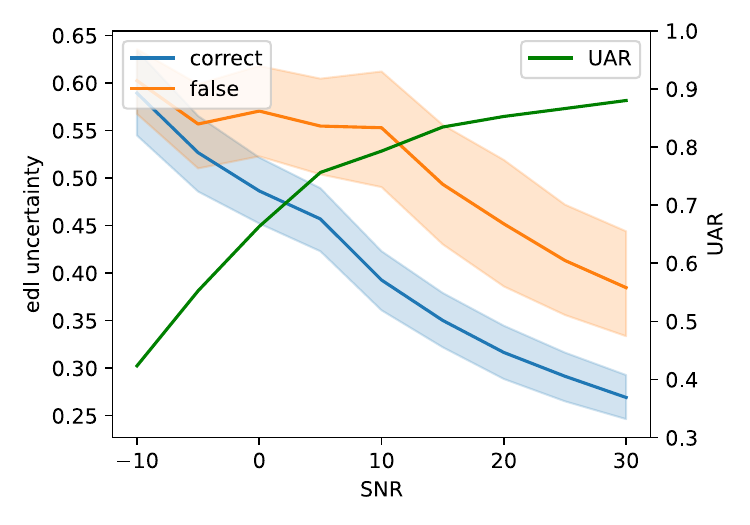}
    \caption*{Evidential DL}
    \end{subfigure}
    \begin{subfigure}[b]{0.19\textwidth}
    \includegraphics[width= \linewidth]{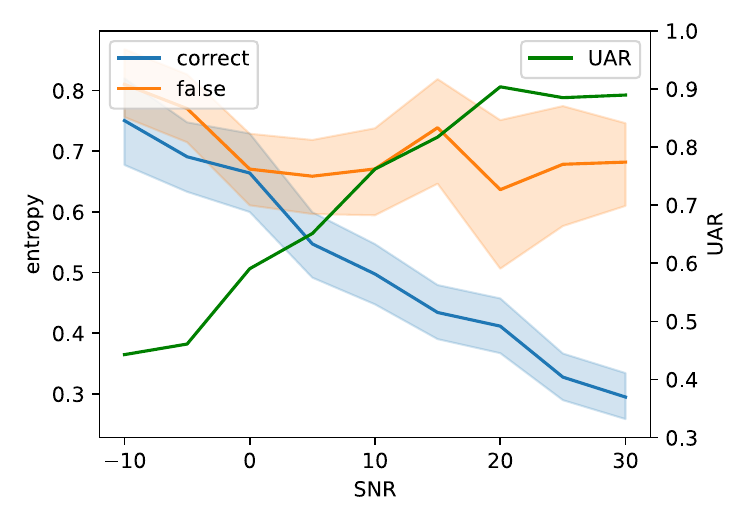}
    \caption*{Prior (in)}
    \end{subfigure}
    \begin{subfigure}[b]{0.19\textwidth}
    \includegraphics[width= \linewidth]{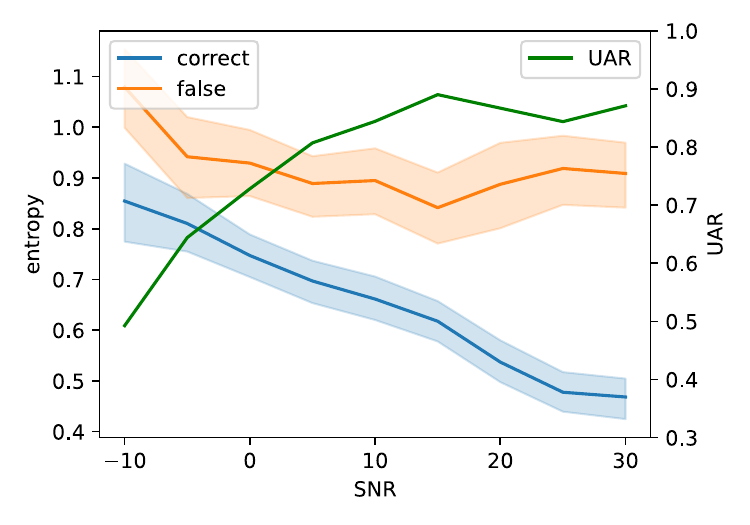}
    \caption*{Prior (out)}
    \end{subfigure}
    \caption{
    Overview of the performance of the 5 \ac{uq} methods on different tests. Higher values on the $x$-axis (and $y$-axis in the bottom row) represent higher uncertainty in all cases.
    The first row shows \ac{CDF} plots for correct and wrong predictions on the test data of all 3 \ac{SER} datasets. 
    The middle row shows \ac{CDF} plots of the uncertainty on speech (blue) and non-speech datasets (red + grey). 
    The bottom row shows the mean uncertainty for EmoDB test data augmented with noise for different \ac{SNR} levels with 95\% confidence interval (left axis) and \ac{UAR} (right axis) for different \ac{SNR}s.}
    \label{fig:overview}
\end{figure*}



\subsection{Model Architecture}

We run the experiments for a \ac{SOTA} wav2vec2.0 pre-trained model for \ac{SER} as described in Wagner et al.\ \cite{wagner2023dawn} followed by a classification head.
We use the first 11 transformer layers of the pre-trained `facebook/wav2vec2-large-robust' model \cite{hsu2021robust} followed by 2 dense layers with a $tanh$ activation function as classification head, which adds up to 153M parameters.

\subsection{Datasets}

We combine three datasets for categorical emotional speech, the two English datasets
\ac{cremad} \cite{cao2014crema},
MSP-Podcast \cite{lotfian2017building}
 (displaying natural emotional expressions),  
 and the
\ac{emodb} \cite{burkhardt2005database} for German acted emotional speech.
All the corpora provide a rater agreement per utterance.
To train a common model for all datasets we choose the subset of the ``basic-4" emotions (anger, happiness, sadness, and neutral) for all datasets, as is also done in \cite{wu2022estimating}\cite{wagner2023dawn}.
Excluding the samples with other emotions, the train set is around 45 h long and contains 28\,945 utterances in total, divided into 60\% neutral, 26\% happiness, 8\% anger, and 6\% sadness.
Beyond, this allows us to treat left out and less frequently occurring emotions as 
\ac{OOD} data.
To test the models when confronted with non-speech data set, we pick cochlscene \cite{jeong2022cochlscene} which contains over 200\,h environmental sounds from 13 different acoustic scenes and the over 100\,h long MUsic, Speech, And Noise Corpus (MUSAN) \cite{snyder2015musan}.


\subsection{Training Procedure}

We train four different models to evaluate the different \ac{uq} methods
without any further fine-tuning or adaptation
\footnote{https://github.com/audeering/ser-uncertainty-quantification}:
The first model, trained with standard 
weighted cross-entropy loss is used for the two baselines entropy and \ac{MC} Dropout \ac{uq}.
For \ac{MC} Dropout, we run 20 forward passes during inference.
The second model is trained with \ac{EDL} loss\footnote{https://github.com/dougbrion/pytorch-classification-uncertainty/} in order to test the corresponding \ac{EDL} \ac{uq}.
Finally, we train two models for \ac{PN}s \cite{malinin2018predictive} with the implementation of Rylan Schaeffer et al.\footnote{ https://github.com/RylanSchaeffer/HarvardAM207-prior-networks}, the first one being only trained on the three \ac{SER}  datasets (PN(in)), while the second one also includes 
cochlscene as \ac{OOD} training data (PN(out)). 

We train all models with a learning rate of 1e-4 and without hyperparameter tuning. It takes around 5 epochs and around 2\,h on an Nvidia RTX A4000 GPU 
for the model to converge.
We use the default, speaker-independent, train, development, and test splits of the datasets.

\subsection{Correctness}\label{sec:corr}

The performances on the test sets are listed in Table \ref{tab:acc}.
For \ac{MC} Dropout, we use the mean over all forward passes.
All models have comparable performance.
\begin{table}[ht]
    \centering
        \caption{\acf{UAR} and Acc on test sets of E = \ac{emodb}, C = \ac{cremad}, M = MSP test 1, CE = Cross Entropy, MC = Monte Carlo Dropout, EDL= Evidential Deep Learning, PN(in and out) for Prior Networks trained on speech (in) and with additional OOD data (out) }
    \label{tab:acc}
    \begin{tabular}{cc|ccccc}
         & & CE & MC & EDL  & PN(in)  & PN(out) \\
        \midrule
        E & Acc & .875 & .875 & .875 & .919 & .904 \\
         &  UAR & .843 &  .843 & .843 &  .898 & .880 \\ 
        \midrule
        C & Acc  & .715 & .728 & .792 & .828 & .766\\
         & UAR & .677 & .704 & .665 & .745 & .711\\
         \midrule
         M & Acc  & .635 & .639 & .662 & .668 & .691\\
         & UAR & .609 & .603 & .576 & .561 & .558\\
    \end{tabular}

\end{table}

In order to test, whether higher uncertainty correlates with a larger number of misclassifications, we 
visualise the uncertainty discrepancy between correct and false predictions with the help of \ac{CDF} plots, similar to 
\cite{sensoy2020uncertainty} (see first row of Figure \ref{fig:overview}).
Ideally, we would expect a considerably larger \ac{AUC} for the correct predictions versus the false predictions in the \ac{CDF} graph.
While to a certain extent true for all \ac{uq}, in particular, our baselines CE and \ac{MC} show very small differences, thus apparently limiting their reliability to detect \ac{ID} misclassifications.

\subsection{Correlation with rater agreement}
We further determine the \ac{PCC} between the \ac{uq} predicitons for all models and the utterance-level rater agreement over all three \ac{SER} datasets. The latter can be calculated as the ratio percentage of raters that chose the majority class as their annotation 
(see Table \ref{tab:agreement}). 


\begin{table}[ht]
    \centering
    \caption{PCC: Correlation between rater agreement and uncertainty, lower is better}
    \label{tab:agreement}
    \begin{tabular}{c|ccccc}
          & CE & MC & EDL  & PN(in)  & PN(out) \\
          \midrule
        EmoDB &  -.236 & -.165  & -.153  & -.186 & -.259 \\
        Crema-d & -.120 & -.174 & -.265  & -.217 & -.238 \\
        MSP-1 &  -.128 & -.082 & -.119 & -.144 & -.194 \\
    \end{tabular}

\end{table}
Even though a weak negative correlation can be observed in all cases, indicating that lower rater agreement also leads to higher uncertainty, the effect is minimal.
A reason for the weak correlation might be that the model only has to distinguish between 4 emotions, while the raters had a larger amount of emotions to discriminate. 
Beyond, even samples with very high annotator disagreement are presented with a “ground truth” emotion during training, thus potentially not sensitising the model for rater disagreements.

\subsection{Out-of-Distribution Emotions}

Table \ref{tab:oodemo} shows the mean uncertainty for emotions included during training vs the emotions omitted during training on the \ac{emodb} test data.
As expected, all methods have greater uncertainty for unseen emotions than for known ones.
\begin{table}[ht]
    \centering
    \caption{Mean Uncertainty for \ac{emodb} emotions included in the training (anger, happiness, neutral, sadness) vs boredom, disgust, fear (out)}
    \label{tab:oodemo}
    \begin{tabular}{c|ccccc}
          & CE & MC & EDL  & PN(in)  & PN(out) \\
          \midrule
        in &  .333 & .010 & .256   &    .239 & .374 \\
        out & .850 & .023 & .387  & .725 & .813 \\
    \end{tabular}

\end{table}

\subsection{Out-of-Domain Data}

In order to test how the \ac{uq} behave under non-speech \ac{OOD} data, we compare the \ac{CDF} plots of the \ac{SER} datasets to those of the other audio databases cochlscene (acoustic ambient noise), musan (speech/music/noise), and additionally, artificially generated ``white noise" (45 samples from near silence with gradually increasing Gaussian white noise) in the second row of Figure \ref{fig:overview}.
As the latter type of data is completely unfit for the \ac{SER} task, we expect lower \ac{AUC} for their graphs (red/grey) compared to those of the speech datasets (blue).
At first glance, it stands out that \ac{PN}(out) has by far the largest separation between \ac{ID} and \ac{OOD} data, as the speech and noise datasets are clearly separated in the \ac{CDF} plot.
A small separation can also be observed for EDL, while any of the other models' \acp{uq} fail to produce a visible distinction. 
It is important to note at this point that the \ac{PN}(out) model is the only one that has been trained on some \ac{OOD} due to its design allowing for this in a straightforward manner.
However, it has only seen \ac{OOD} data in the form of ambient noise from cochlscene, and it manages to generalise well to other types of \ac{OOD} data.
Regarding the individual databases, EmoDB has almost always the lowest uncertainty, which aligns with models showing the highest performance on it.

\subsection{Sensor Uncertainty}
Our final test aims to simulate sensor uncertainty by augmenting the \ac{emodb} dataset with different levels of white noise, starting with an \acf{SNR} of 30 (original utterance is 30\,dB louder than the noise) and increasing noise levels up to an \ac{SNR} of -10.
The last row of Figure \ref{fig:overview} shows the mean uncertainty for correct and false predictions over the different \ac{SNR}s.
The plot also contains the \ac{UAR}, which, as expected, gets progressively worse as the noise level increases.
For an SNR of 30\,dB, the uncertainties for the correct and false predictions are clearly separated for all \ac{uq} methods, even more prominently than in the top row, as we only evaluate on the `easiest' dataset \ac{emodb}.

For increasing noise levels, we would expect an increasing uncertainty for the correct predictions as more and more samples are correctly classified by accident, which we can observe to varying degrees for all methods.
However, with (Cross) Entropy and \ac{MC} Dropout, the uncertainty of the misclassified samples also decreases, which makes it impossible to reliably distinguish between false and correct predictions based on the \ac{uq} at this point.
For EDL however, we see exactly the opposite behaviour, as all predictions become more uncertain with decreasing \ac{SNR}, which is a promising behaviour in order to prevent faulty classifications based on \ac{uq}.
Still, the fact that the uncertainty increases much quicker than \ac{UAR} indicates that the \ac{uq} is too sensitive to noise.
Interestingly, for EDL we see higher uncertainties for speech samples with added noise than for pure white noise in the middle row of Figure \ref{fig:overview}.
The most ideal behaviour is shown by the two \ac{PN}s with the uncertainty of false predictions initially remaining stable, but the uncertainty of correct predictions steadily increasing.

\section{Conclusions and Outlook}\label{sec:conclusion}
We tested the behaviour of different \ac{uq} methods under realistic sources of uncertainty for \ac{SER} applications, such as \ac{OOD} emotions, \ac{OOD} data or corrupted signals.  
We find that entropy-based \ac{uq} predictions on models with a standard cross-entropy training pipeline can already give good intuition on whether an emotion prediction can be considered trustworthy, thus reducing the importance of additional gatekeepers such as SNR models or voice activity detection systems.
Nevertheless, the reliability of \ac{uq} can greatly be increased by incorporating the \ac{uq} already in the network design, in particular through Prior Networks and by exposing the model to \ac{OOD} data during training. 
In this regard, further improvement might be expected through methods that artificially create the \ac{OOD} data \cite{van2020uncertainty} or calibration \cite{naeini2015obtaining}.
However, a closer investigation about the separability of data uncertainty and \ac{OOD} data through \ac{PN}s and other \acp{uq} in the context of \ac{SER} seems necessary.

\section{Acknowledgements} 

Manuel Milling and  Björn Schuller are with MCML -- Munich Center for Machine Learning, Germany,  Björn Schuller is with MDSI -- Munich Data Science Institute, Germany

This research has been partly funded by the European SHIFT (MetamorphoSis of cultural Heritage Into augmented hypermedia assets For enhanced accessibiliTy and inclusion) project (Grant Agreement number: 101060660).


\bibliographystyle{IEEEtran}
\bibliography{main}

\begin{thebibliography}{10}
\providecommand{\url}[1]{#1}
\csname url@samestyle\endcsname
\providecommand{\newblock}{\relax}
\providecommand{\bibinfo}[2]{#2}
\providecommand{\BIBentrySTDinterwordspacing}{\spaceskip=0pt\relax}
\providecommand{\BIBentryALTinterwordstretchfactor}{4}
\providecommand{\BIBentryALTinterwordspacing}{\spaceskip=\fontdimen2\font plus
\BIBentryALTinterwordstretchfactor\fontdimen3\font minus \fontdimen4\font\relax}
\providecommand{\BIBforeignlanguage}[2]{{%
\expandafter\ifx\csname l@#1\endcsname\relax
\typeout{** WARNING: IEEEtran.bst: No hyphenation pattern has been}%
\typeout{** loaded for the language `#1'. Using the pattern for}%
\typeout{** the default language instead.}%
\else
\language=\csname l@#1\endcsname
\fi
#2}}
\providecommand{\BIBdecl}{\relax}
\BIBdecl

\bibitem{schuller2018speech}
B.~W. Schuller, ``Speech emotion recognition: Two decades in a nutshell, benchmarks, and ongoing trends,'' \emph{Communications of the ACM}, vol.~61, no.~5, pp. 90--99, 2018.

\bibitem{sahu2019multimodal}
G.~Sahu, ``Multimodal speech emotion recognition and ambiguity resolution,'' \emph{arXiv e-prints}, pp. arXiv--1904, 2019.

\bibitem{milling2022evaluating}
M.~Milling, A.~Baird, K.~D. Bartl-Pokorny, S.~Liu, A.~M. Alcorn, J.~Shen, T.~Tavassoli, E.~Ainger, E.~Pellicano, M.~Pantic \emph{et~al.}, ``Evaluating the impact of voice activity detection on speech emotion recognition for autistic children,'' \emph{Frontiers in Computer Science}, vol.~4, p. 837269, 2022.

\bibitem{rizos2020average}
G.~Rizos and B.~W. Schuller, ``Average jane, where art thou?--recent avenues in efficient machine learning under subjectivity uncertainty,'' in \emph{Information Processing and Management of Uncertainty in Knowledge-Based Systems: 18th International Conference, IPMU 2020, Lisbon, Portugal, June 15--19, 2020, Proceedings, Part I 18}.\hskip 1em plus 0.5em minus 0.4em\relax Springer, 2020, pp. 42--55.

\bibitem{gerczuk2021emonet}
M.~Gerczuk, S.~Amiriparian, S.~Ottl, and B.~W. Schuller, ``Emonet: A transfer learning framework for multi-corpus speech emotion recognition,'' \emph{IEEE Transactions on Affective Computing}, vol.~14, no.~2, pp. 1472--1487, 2021.

\bibitem{jaiswal2021best}
M.~Jaiswal and E.~Mower~Provost, ``Best practices for noise-based augmentation to improve the performance of emotion recognition" in the wild",'' \emph{arXiv e-prints}, pp. arXiv--2104, 2021.

\bibitem{abdar2021review}
M.~Abdar, F.~Pourpanah, S.~Hussain, D.~Rezazadegan, L.~Liu, M.~Ghavamzadeh, P.~Fieguth, X.~Cao, A.~Khosravi, U.~R. Acharya \emph{et~al.}, ``A review of uncertainty quantification in deep learning: Techniques, applications and challenges,'' \emph{Information Fusion}, vol.~76, pp. 243--297, 2021.

\bibitem{gawlikowski2023survey}
J.~Gawlikowski, C.~R.~N. Tassi, M.~Ali, J.~Lee, M.~Humt, J.~Feng, A.~Kruspe, R.~Triebel, P.~Jung, R.~Roscher \emph{et~al.}, ``A survey of uncertainty in deep neural networks,'' \emph{Artificial Intelligence Review}, vol.~56, no. Suppl 1, pp. 1513--1589, 2023.

\bibitem{hullermeier2021aleatoric}
E.~H{\"u}llermeier and W.~Waegeman, ``Aleatoric and epistemic uncertainty in machine learning: An introduction to concepts and methods,'' \emph{Machine Learning}, vol. 110, pp. 457--506, 2021.

\bibitem{tan2023single}
A.~R. Tan, S.~Urata, S.~Goldman, J.~C. Dietschreit, and R.~G{\'o}mez-Bombarelli, ``Single-model uncertainty quantification in neural network potentials does not consistently outperform model ensembles,'' \emph{npj Computational Materials}, vol.~9, no.~1, p. 225, 2023.

\bibitem{liu2020simple}
J.~Liu, Z.~Lin, S.~Padhy, D.~Tran, T.~Bedrax~Weiss, and B.~Lakshminarayanan, ``Simple and principled uncertainty estimation with deterministic deep learning via distance awareness,'' \emph{Advances in Neural Information Processing Systems}, vol.~33, pp. 7498--7512, 2020.

\bibitem{zhang2013active}
Z.~Zhang, J.~Deng, E.~Marchi, and B.~Schuller, ``Active learning by label uncertainty for acoustic emotion recognition,'' in \emph{Proceedings INTERSPEECH 2013, 14th Annual Conference of the International Speech Communication Association, Lyon, France}, 2013.

\bibitem{han2017hard}
J.~Han, Z.~Zhang, M.~Schmitt, M.~Pantic, and B.~Schuller, ``From hard to soft: Towards more human-like emotion recognition by modelling the perception uncertainty,'' in \emph{Proceedings of the 25th ACM international conference on Multimedia}, 2017, pp. 890--897.

\bibitem{amir2015comparing}
N.~Amir, R.~Rubinstein, A.~Shlomov, and G.~Diamond, ``Comparing categorical and dimensional ratings of emotional speech,'' in \emph{2015 11th IEEE International Conference and Workshops on Automatic Face and Gesture Recognition (FG)}, vol.~5.\hskip 1em plus 0.5em minus 0.4em\relax IEEE, 2015, pp. 1--5.

\bibitem{shannon1948mathematical}
C.~E. Shannon, ``A mathematical theory of communication,'' \emph{The Bell system technical journal}, vol.~27, no.~3, pp. 379--423, 1948.

\bibitem{kong2020panns}
Q.~Kong, Y.~Cao, T.~Iqbal, Y.~Wang, W.~Wang, and M.~D. Plumbley, ``Panns: Large-scale pretrained audio neural networks for audio pattern recognition,'' \emph{IEEE/ACM Transactions on Audio, Speech, and Language Processing}, vol.~28, pp. 2880--2894, 2020.

\bibitem{baevski2020wav2vec}
A.~Baevski, Y.~Zhou, A.~Mohamed, and M.~Auli, ``wav2vec 2.0: A framework for self-supervised learning of speech representations,'' \emph{Advances in neural information processing systems}, vol.~33, pp. 12\,449--12\,460, 2020.

\bibitem{gal2016dropout}
Y.~Gal and Z.~Ghahramani, ``Dropout as a bayesian approximation: Representing model uncertainty in deep learning,'' in \emph{international conference on machine learning}.\hskip 1em plus 0.5em minus 0.4em\relax PMLR, 2016, pp. 1050--1059.

\bibitem{sensoy2018evidential}
M.~Sensoy, L.~Kaplan, and M.~Kandemir, ``Evidential deep learning to quantify classification uncertainty,'' \emph{Advances in neural information processing systems}, vol.~31, 2018.

\bibitem{Amini_2020}
\BIBentryALTinterwordspacing
A.~Amini, W.~Schwarting, A.~Soleimany, and D.~Rus, ``Deep evidential regression,'' in \emph{Advances in Neural Information Processing Systems}, H.~Larochelle, M.~Ranzato, R.~Hadsell, M.~Balcan, and H.~Lin, Eds., vol.~33.\hskip 1em plus 0.5em minus 0.4em\relax Curran Associates, Inc., 2020, pp. 14\,927--14\,937. [Online]. Available: \url{https://proceedings.neurips.cc/paper/2020/file/aab085461de182608ee9f607f3f7d18f-Paper.pdf}
\BIBentrySTDinterwordspacing

\bibitem{liu2008classic}
L.~Liu and R.~R. Yager, ``Classic works of the dempster-shafer theory of belief functions: An introduction,'' \emph{Classic works of the Dempster-Shafer theory of belief functions}, pp. 1--34, 2008.

\bibitem{malinin2018predictive}
A.~Malinin and M.~Gales, ``Predictive uncertainty estimation via prior networks,'' \emph{Advances in neural information processing systems}, vol.~31, 2018.

\bibitem{wagner2023dawn}
J.~Wagner, A.~Triantafyllopoulos, H.~Wierstorf, M.~Schmitt, F.~Burkhardt, F.~Eyben, and B.~W. Schuller, ``Dawn of the transformer era in speech emotion recognition: closing the valence gap,'' \emph{IEEE Transactions on Pattern Analysis and Machine Intelligence}, 2023.

\bibitem{hsu2021robust}
W.-N. Hsu, A.~Sriram, A.~Baevski, T.~Likhomanenko, Q.~Xu, V.~Pratap, J.~Kahn, A.~Lee, R.~Collobert, G.~Synnaeve \emph{et~al.}, ``Robust wav2vec 2.0: Analyzing domain shift in self-supervised pre-training,'' \emph{arXiv e-prints}, pp. arXiv--2104, 2021.

\bibitem{cao2014crema}
H.~Cao, D.~G. Cooper, M.~K. Keutmann, R.~C. Gur, A.~Nenkova, and R.~Verma, ``Crema-d: Crowd-sourced emotional multimodal actors dataset,'' \emph{IEEE transactions on affective computing}, vol.~5, no.~4, pp. 377--390, 2014.

\bibitem{lotfian2017building}
R.~Lotfian and C.~Busso, ``Building naturalistic emotionally balanced speech corpus by retrieving emotional speech from existing podcast recordings,'' \emph{IEEE Transactions on Affective Computing}, vol.~10, no.~4, pp. 471--483, 2017.

\bibitem{burkhardt2005database}
F.~Burkhardt, A.~Paeschke, M.~Rolfes, W.~F. Sendlmeier, B.~Weiss \emph{et~al.}, ``A database of german emotional speech.'' in \emph{Interspeech}, vol.~5, 2005, pp. 1517--1520.

\bibitem{wu2022estimating}
W.~Wu, C.~Zhang, X.~Wu, and P.~C. Woodland, ``Estimating the uncertainty in emotion class labels with utterance-specific dirichlet priors,'' \emph{IEEE Transactions on Affective Computing}, 2022.

\bibitem{jeong2022cochlscene}
I.-Y. Jeong and J.~Park, ``Cochlscene: Acquisition of acoustic scene data using crowdsourcing,'' in \emph{2022 Asia-Pacific Signal and Information Processing Association Annual Summit and Conference (APSIPA ASC)}.\hskip 1em plus 0.5em minus 0.4em\relax IEEE, 2022, pp. 17--21.

\bibitem{snyder2015musan}
D.~Snyder, G.~Chen, and D.~Povey, ``Musan: A music, speech, and noise corpus,'' \emph{arXiv e-prints}, pp. arXiv--1510, 2015.

\bibitem{sensoy2020uncertainty}
M.~Sensoy, L.~Kaplan, F.~Cerutti, and M.~Saleki, ``Uncertainty-aware deep classifiers using generative models,'' in \emph{Proceedings of the AAAI Conference on Artificial Intelligence}, vol.~34, no.~04, 2020, pp. 5620--5627.

\bibitem{van2020uncertainty}
J.~Van~Amersfoort, L.~Smith, Y.~W. Teh, and Y.~Gal, ``Uncertainty estimation using a single deep deterministic neural network,'' in \emph{International conference on machine learning}.\hskip 1em plus 0.5em minus 0.4em\relax PMLR, 2020, pp. 9690--9700.

\bibitem{naeini2015obtaining}
M.~P. Naeini, G.~Cooper, and M.~Hauskrecht, ``Obtaining well calibrated probabilities using bayesian binning,'' in \emph{Proceedings of the AAAI conference on artificial intelligence}, vol.~29, no.~1, 2015.

\end{thebibliography}

\end{document}